\documentclass[12pt,preprint]{aastex}

\newcommand{\msun}{M$_\sun$}

\newcommand{\om}{$\Omega_m$}

\slugcomment{Last modified \today}

\begin{document}

\title{Evolution of the Cluster Correlation Function}

\author{
Neta A. Bahcall\altaffilmark{1},
Lei Hao\altaffilmark{1},
Paul Bode\altaffilmark{1},
Feng Dong\altaffilmark{1}
}

\altaffiltext{1}{Princeton University Observatory, Princeton, NJ 08544}

\begin{abstract}

We study the evolution of the cluster correlation function 
and its richness-dependence from $z = 0$ to $z = 3$ using large-scale 
cosmological simulations. A standard flat LCDM model with 
$\Omega_m = 0.3$ and, for comparison, a tilted $\Omega_m = 1$ model, TSCDM, 
are used. The evolutionary predictions are presented in a format 
suitable for direct comparisons with observations. We find that the 
cluster correlation strength increases with redshift: high redshift 
clusters are clustered more strongly (in comoving scale) than low 
redshift clusters of the same mass. The increased correlations with 
redshift, in spite of the decreasing mass correlation strength, is 
caused by the strong increase in cluster bias with redshift: 
clusters represent higher density peaks of the mass distribution as 
the redshift increases. The richness-dependent cluster correlation 
function, presented as the correlation-scale versus cluster mean 
separation relation, $R_0$ - $d$, is found to be, remarkably, independent 
of redshift to $z \la 2$ for LCDM and $z \la 1$ for TCDM (for a fixed 
correlation function slope and cluster mass within a fixed comoving radius). 
The non-evolving $R_0$ - $d$ relation 
implies that both the comoving clustering scale and the cluster 
mean separation increase with redshift for the same mass clusters 
so that the $R_0$ - $d$ relation remains essentially unchanged. For LCDM, this 
relation is $R_0(z) \simeq 2.6\sqrt{d(z)}$ for $z \la 2$ 
(in comoving $h^{-1}$ Mpc scales). TSCDM has smaller correlation 
scales, as expected. Evolution in the relation is seen at $z \ga 2$ 
for LCDM and $z \ga 1$ for TSCDM, 
where the amplitude of the relations declines. The evolution of the 
$R_0$ - $d$ relation from $z \sim 0$ to $z \sim 3$ provides an important 
new tool in cosmology; it can be used to break degeneracies that exist 
at $z \sim 0$ and provide precise determination of cosmological parameters. 

\end{abstract}


\section{Introduction} \label{introduction}

The correlation function of clusters of galaxies provides a 
powerful test of cosmological models: both the amplitude of 
the correlation function and its dependence on cluster mass/richness 
are determined by the underlying cosmology. It has long been shown 
that clusters are more strongly correlated in space than galaxies --  
by an order-of-magnitude: the typical galaxy correlation scale of 
$\sim$ 5 $h^{-1}$ Mpc increases to $\sim$ 20 - 25 $h^{-1}$ Mpc for the 
richest clusters (\citealt{bah83, kly83}, see also \citealt{bah88, huc90, 
pos92, bah92, pea92, dal94, cro97, aba98, lee99, bor99, col00, gon02, bah03c}; 
and references therein). \citet{bah83} showed that the cluster correlation 
function is richness dependent: the correlation strength increases 
with cluster richness, or mass. The rarest, most massive clusters exhibit the 
strongest spatial correlations. Many observations have since confirmed these 
results (see references above), and theory has nicely explained them 
(\citealt{kai84, bah92a, moh96, gov99, colb00, mos01, bah03c}; and references 
therein). All analyses so far have been done at small redshifts, $z \la 0.5$.
As cluster samples become available at larger redshifts, it is important to 
determine the expected evolution of the cluster correlation function and the 
evolution of its richness dependence as a function of redshift; these will
provide direct cosmological predictions for comparisons with the data. Analytic 
approximations for the evolution of cluster halo abundance, bias and clustering have 
been presented by several groups \citep{man93, moh96, moh02, she01, mos01}. 
Here we use direct N-body cosmological simulations to determine the
expected evolution of the cluster correlation function and present the results 
in a format suitable for direct comparison with observations.

We use large-scale high-resolution cosmological simulations to investigate the 
evolution of the cluster correlation function from $z = 0$ to $z = 3$ over 
a wide range of cluster masses. We determine the evolution of the richness dependence 
of the cluster correlation function over the same redshift range. The simulations use 
a standard flat LCDM model which best fits numerous other observations (e.g., \citealt{bah99, 
ben03, spe03}). For comparison, we also investigate the evolution of the cluster correlation 
function in an $\Omega_m = 1$ tilted SCDM model.

\section{Simulations} \label{simulation}

The high-resolution N-body simulations use a standard flat Lambda CDM cosmology 
with $\Omega_m = 0.3$, $h = 0.67$, $n = 1$, and $\sigma_8 = 0.9$. For comparison, 
we also investigate an $\Omega_m = 1$ Tilted SCDM model (TSCDM; with $h = 0.5$, $\sigma_8 = 0.5$, 
$n = 0.625$). The simulations are discussed in detail by \citet{bod00, bod01}. 
A brief summary is given below.

The simulations employ $512^3 = 1.34\times10^8$ particles in a periodic 
cube, with a particle mass of $6.2\times10^{11} h^{-1}$ \msun. A larger simulation with 
$1024^3$ particles and one eighth the particle mass was also investigated for comparison; 
it yields consistent results. 
The box length is 1000 $h^{-1}$ Mpc for the LCDM model (669.4 $h^{-1}$ Mpc for TSCDM). 
The spline kernel softening length is 27 $h^{-1}$ kpc. The spatial resolution is 
comfortably smaller than the typical $\sim$100 kpc core size of clusters. 
The N-body evolution was carried out with the tree-particle-mesh (TPM) code \citep{bod00, bod03}.
These are among the largest volume simulations currently available.

Clusters were selected in the simulations using the HOP algorithm for the detection of 
high-density peaks \citep{eis98} as described in \citet{bod01}. 
Cluster masses are defined within a comoving radius of 1.5 $h^{-1}$ Mpc of the cluster center. 
Cluster masses within other fixed radii yield similar results. Clusters with mass threshold 
of M$_{1.5} \geq 2\times10^{13} \ h^{-1}$ \msun \ (within comoving radius 1.5 $h^{-1}$ Mpc) 
are selected at redshifts $z$ = 0, 0.17, 0.5, 1, 2, and 3. The number of clusters used
ranges from $2\times10^5$ clusters (for $z$ = 0, M$_{1.5} \geq 2\times10^{13}$ \msun) 
to $\sim 10^3$ clusters for the highest mass, highest redshift clusters. 
The abundance of LCDM clusters ($2\times10^{-4}$ and $1.4\times10^{-5} \ h^3$ Mpc$^{-3}$ for 
M$_{1.5} \geq 2\times10^{13}$ and $2\times10^{14} \ h^{-1}$ \msun \ at $z$ = 0) is consistent 
with recent observations of SDSS and X-ray clusters \citep{bah03a, ike02, rei02}. 
For more details see \citet{bod01}.

\section{Evolution of the Cluster Correlation Function} \label{evolution}

We determine the two-point spatial correlation function of clusters at different 
redshifts as a function of cluster mass threshold. The comoving mass thresholds range 
from M$_{1.5} \geq 2\times10^{13} \ h^{-1}$ \msun \ (within 1.5 $h^{-1}$ comoving Mpc) to 
$5\times10^{14} \ h^{-1}$ \msun. The redshifts investigated range from $z$ = 0 to 3. 
The correlation function is determined for each sample by comparing the observed 
distribution of cluster pairs as a function of pair separation with the distribution 
in random catalogs within the same volume: $\xi_{cc}(r) = F_{DD}(r)/F_{RR}(r) - 1$, 
where $F_{DD}$($r$) and $F_{RR}$($r$) are the frequencies of cluster-cluster pairs as 
a function of pair separation $r$ in the data and in random catalogs, respectively. 
The number of clusters decreases with increasing mass and redshift, from $2\times10^5$ to 
$\sim10^3$ clusters; the most massive (rarest) clusters are therefore studied only at the 
lower part of the redshift range. Since the cluster abundance in a TSCDM model decreases 
more rapidly with redshift than in LCDM, this model is only studied up to $z = 1$. Poisson 
statistical error-bars are used in the correlation function analysis. Comoving scales 
and a Hubble constant of H$_0 = 100 \ h \ km \ s^{-1}$ Mpc$^{-1}$ 
are used throughout.

The evolution of the cluster correlation function as a function of redshift is 
illustrated in Figure 1 for LCDM for two mass threshold samples ($2\times10^{13}$ and 
$1\times10^{14} \ h^{-1}$ \msun). The best-fit power-law correlation function, 
$\xi$($r$) = ($\frac{r}{R_{0}}$)$^{-\gamma}$, is also presented for each sample. 
The best fits are derived for scales $r \leq 50 \ h^{-1}$ Mpc. The power-law slope was 
treated both as a free parameter and as a fixed value of $\gamma = 2$; the latter is 
the typical slope found in the simulations. The best-fits shown in Figure 1 are 
for a fixed slope $\gamma = 2$. The results are similar for a free slope fit, as discussed 
below. The evolution of the cluster correlation function is apparent in Figure 1: 
clusters are more strongly correlated at higher redshift. This is of course opposite to the 
evolution of the mass correlation function, which decreases with redshift.
The enhancement of the cluster correlation strength with redshift is due to 
the increased bias of the clusters relative to the underlying mass distribution: 
i.e., the same comoving mass clusters represent higher density peaks of the mass 
distribution as the redshift increases thus amplifying their correlation strength 
(see, for example, \citealt{col89, moh96, moh02, she01, mos01})

Figure 2 presents the best-fit correlation function slope $\gamma$ as a function 
of cluster redshift and cluster mass. The best slope for LCDM is $\gamma \sim 2$ for 
$z \la 0.5$ clusters. The slope steepens by $\sim 20\%$, to $\gamma \sim 2.3 - 2.5$, 
as the redshift increases to $z \sim 2 - 3$. This steepening can also be seen in 
the correlation function of the high-redshift samples in Figure 1. The TSCDM model (Figure 2b) 
yields a slightly steeper average slope: $\sim$ 2.3 at $z \la 0.5$, increasing only 
slightly to $\sim 2.5$ at $z \sim 1$.

The evolution of the cluster correlation function is presented in Figure 3 for clusters 
with different mass thresholds. The figure illustrates the dependence of the comoving correlation 
scale, $R_0$, on redshift. The results are shown for a correlation slope 
of 2 as well as for a free fit slope. As expected, the slight steepening of the best-fit 
slope at high redshift causes the correlation scale at high redshift to be somewhat 
smaller than for $\gamma = 2$. The changing slope does not change the main 
evolutionary trend of $R_0$($z$). Two main results are apparent in Figure 3: 
a) the cluster correlation scale increases with redshift; b) the evolutionary increase of 
the correlation scale is stronger for the more massive clusters as well as 
at higher redshifts; low mass and low redshift clusters show only a small increase of $R_0$ 
with $z$. For example, the low mass sample with M$_{1.5} \geq 2\times10^{13}$ \msun \ 
increases its correlation scale from $\sim 10 \ h^{-1}$ Mpc at $z = 0$ to $11 \ h^{-1}$ Mpc at $z = 1$ 
and $18 \ h^{-1}$ Mpc at $z = 3$ (for LCDM), while M$_{1.5} \geq 2\times10^{14}$ \msun \ clusters 
increase their $R_0$ from $\sim 16 \ h^{-1}$ Mpc at $z = 0$ to $25 \ h^{-1}$ Mpc at $z = 1$. 
The TSCDM model (Figure 3b) shows a similar (but faster) increase of $R_0$ with $z$. 
As expected, the $R_0$ values of the TSCDM model are smaller than those of LCDM (at low redshift). 
The observed increase of the correlation scale with cluster mass threshold seen in Figure 3 
is the well-known richness-dependence of the cluster correlation function \citep{bah83, 
pea92, bah92, moh96, moh02, gov99, col00, bah03c}. The strengthening of the cluster correlations 
with redshift implies that the cluster bias increases with redshift more strongly than the 
mass correlations decrease with redshift.

\section{Evolution of the Ro - d Relation} \label{relation}

The cluster correlation stength increases with both cluster mass and cluster redshift. 
Here we combine the two by investigating the evolution of the richness-dependent 
cluster correlation function. We present the richness-dependent correlation function as the 
dependence of the correlation scale $R_0$ on the cluster mean separation $d$ (\citealt{bah83, 
sza85, bah88, cro97, gov99, colb00, bah03c}; and references therein). Samples with 
intrinsically larger mean separation correspond to lower intrinsic cluster abundance 
($n_{cl} = d^{-3}$) and thus to higher cluster richness and mass (for complete samples). 
Since the cluster mass does not enter this relation, the predictions can be easily compared 
with observations.

In Figure 4 we present the $R_0 - d$ relation for different redshifts for the 
LCDM and TSCDM models. A fixed correlation function slope of $\gamma = 2$ is used. 
Remarkably, the richness-dependent correlation function, $R_0 - d$, is independent of 
redshift for $z \la 2$ for LCDM and $z \la 1$ for TSCDM. The scatter in the $R_0 - d$ 
relation with redshift is small when a fixed slope ($\gamma = 2$) is used in determining 
$R_0$. For a free fit slope (LCDM; Figure 5), the correlation scales at high redshift are 
slightly lower due to the steepening of the slope $\gamma$ at high $z$. This results in a 
small amount of evolution in the $R_0 - d$ relation for $d \ga 50\ h^{-1}$ Mpc, where 
the amplitude of the relation decreases by $\la 10\%$ as the redshift 
increases from $z = 0$ to $z = 2$ (Figure 5; to $d \sim 100 \ h^{-1}$ Mpc). The $R_0 - d$ relation 
for the TSCDM model remains essentially un-evolving for $z < 1$ for both fixed and free slopes (Figure 5). 
The best-fit slope for this model changes only slightly with redshift 
(Figure 2). Using a slope of 2.3 instead of 2 makes only a small difference in the $R_0 - d$ 
relation (Figure 3). The $R_0 - d$ relation evolves at $z \ga 2$ for LCDM ($\gamma = 2$), where 
the amplitude of the relation declines. The same is suggested for TSCDM at $z \ga 1$. 
The above results use cluster masses defined within a fixed comoving radius (1.5 $h^{-1}$ Mpc); 
these masses are observationally easy to determine. Using cluster virial masses (or Friend-of-Friend 
masses to a given mass-density threshold), which are observationally more difficult to determine, 
yield similar results but with slightly more evolution at large $d$'s ($\la 10\%$ decrease in the $R_0 - d$ 
amplitude as the redshift increases from $z = 0$ to 2 for LCDM with $\gamma = 2$, and $\la 15\%$ decrease 
for a free-fit slope). Only little evolution is seen for $z \la 1$ ($< 10\%$). In addition, the 
amplitude of the $R_0 - d$ relation is slightly lower ($\la 10\%$) at small separations 
($d \la 40\ h^{-1}$ Mpc) for these masses, but with no significant evolution to $z \la 1$. 
An analysis of the clustering in \om= 1 mixed dark matter models \citep{gar00} similarly yields an 
$R_0 - d$ relation that is essentially the same at $z = 0$ and $z = 0.8$. 
The results presented here, for cluster masses within a fixed comoving radius, can be directly compared 
with observations; the latter are to be defined in a similar manner. 

A higher resolution LCDM simulation with $1024^3$ particles yields consistent results (within $5\%$ 
for $d \la 60\ h^{-1}$ Mpc and $10\%$ ($\sim$ 2-$\sigma$ level) for $d > 60\ h^{-1}$ Mpc) 
with those presented in Figures 4 and 5.

The remarkable constancy of the $R_0 - d$ relation to these high redshifts, with no significant 
evolution, implies that the increased cluster correlation strength with redshift 
(Figure 3) is matched by the increased mean separation $d$ (i.e., lower cluster abundance) 
so that the $R_0 - d$ relation remains essentially unchanged. For a given mass cluster, the cluster 
abundance decreases with redshift; the decrease is rapid for low $\sigma_8$ models and 
slow for high $\sigma_8$ models (e.g., \citealt{bah98}); the cluster mean separation $d$ 
therefore increases with redshift, as expected. The evolution of the cluster correlation 
scale $R_0$ depends on the combined evolution of the underlying mass correlation function 
(which decreases with increasing redshift: the decrease is rapid for $\Omega_m = 1$ models 
and slower for low $\Omega_m$ models), and the evolution of the cluster bias relative 
to the underlying mass distribution. The cluster bias increases with redshift: the same 
mass clusters represent higher density peaks at high redshift. The bias increases more 
rapidly for high $\Omega_m$ models than for lower $\Omega_m$ models (see, e.g., \citealt{col89, 
moh96, moh02, she01}). The combined effect of the decreasing mass correlation strength and 
the increasing cluster bias with redshift (for same mass clusters) results in an increasing 
cluster correlation scale with redshift as seen in Figure 3. The increased $R_0$($z$) 
matches the increase in the mean separation $d$($z$) in the relative $R_0 - d$ relation (for 
$\gamma = 2$). Since $R_0 \sim d^{0.5}$ (see below), $R_0$($z$) increases roughly as $d$($z$)$^{0.5}$. 
The same mass clusters therefore shift upwards along the $R_0 - d$ relation, to larger $d$ and 
larger $R_0$, as the redshift increases. At some high redshift ($z \sim 2$ for LCDM; $z \sim 1$ 
for TSCDM) the drop in cluster abundance (increase in $d$) becomes considerably stronger than 
the increase in clustering scale $R_0$, and the amplitude of the $R_0 - d$ relation declines as 
seen in Figure 4. The results are not sensitive to the precise value of $\sigma_8$; the 
$R_0 - d$ relation changes by $\la 10\%$ at low redshifts as $\sigma_8$ changes from 0.9 to 0.7 
(values within this range are suggested by recent observations; see, e.g., \citealt{ike02, rei02, 
bah03a, pie03, sch03, spe03, teg03}). The redshift at which evolution in the $R_0 - d$ relation becomes 
significant is expected to depend on $\sigma_8$; the detailed dependance needs to be investigated 
by simulations. 

The evolution of the $R_0 - d$ relation provides an important cosmological tool. The results 
can be used for direct comparison with observations at any redshift ($z \la 3$). No significant evolution in 
the $R_0 - d$ relation is expected to $z \la 2$ (using $\gamma = 2$ and cluster mass within fixed comoving 
radius) if the current LCDM model 
is correct. An approximation for the $R_0 - d$ relation for LCDM is given by 
(see \citealt{bah03c}):
 \begin{equation}
     R_0(z) \simeq  2.6 \times d(z) ^{0.5}  \ \ \ \ \ \ \ \  
(LCDM; \ \ z \la 2; \ \ \gamma =2; \ \ \ d \simeq20-100)
\end{equation}
where all scales are in comoving $h^{-1}$ Mpc. This approximation holds for all $z \la 2$ clusters 
(using a slope $\gamma = 2$ and cluster mass within fixed comoving radius) in the range 
$d \sim 20 - 100 h^{-1}$ Mpc. As cluster samples 
become available at high redshift, this comparison should provide an important test of the 
LCDM cosmology.

Observations of the cluster correlation function at $z \sim 0$ have been influential in 
cosmology; they provided some of the earliest evidence that contradicted the then standard 
$\Omega_m = 1$ model and indicated a low-density universe (e.g. \citealt{whi87, bah88, bah92a}). 
But the $z \sim 0$ data are insufficient by themselves to place a precise constraint on the 
cosmological parameters. This is so not only because of the observational scatter among 
different samples, but mainly because the $z \sim 0$ relation is degenerate with respect to 
different cosmological parameters. For example, the amplitude of the $R_0 - d$ 
relation at $z \sim 0$ increases with increasing $\sigma_8$ but it also increases with 
decreasing $\Omega_m$; models with higher $\sigma_8$ and higher $\Omega_m$ 
yield degenerate results with models of lower $\sigma_8$ and lower $\Omega_m$. 
The slope of the power-spectrum of mass fluctuation introduces yet another free parameter in the degeneracy. 
This degeneracy can be broken by studying the evolution of the $R_0 - d$ relation to high redshifts, 
since the evolution depends differently on the combination of the cosmological 
parameters. While other independent observations can also be used -- 
such as the cluster mass function and its evolution, and the CMB spectrum of fluctuations 
-- the $R_0 - d$ ($z$) evolution provides an independent consistency test that uses a single 
self-consistent method of cluster correlations.

\section{Conclusions}

We use large-scale high-resolution cosmological simulations to determine the evolution 
of the cluster correlation function with redshift from $z = 0$ to $z = 3$ 
over a wide range of cluster masses. Two cosmological models are studied: the 
standard flat LCDM model (with $\Omega_m = 0.3$), which best fits numerous observations, 
and, for comparison, a tilted $\Omega_m = 1$ model (TSCDM). The evolutionary predictions 
are presented in a format suitable for direct comparisons with observations.

We find that the cluster correlation strength increases with 
redshift for fixed mass clusters; i.e., clusters are more strongly 
clustered in space at high redshift. The evolutionary increase 
of the correlation scale with redshift (in comoving units) is faster 
for more massive clusters and at higher redshift. The increased 
clustering of clusters at high redshift, in spite of the decreased 
clustering of the underlying mass distribution, is due to the 
strongly increasing bias of clusters at high redshift: clusters 
represent higher-density peaks of the mass distribution at higher 
redshift. The increased bias dominates over the decreasing mass 
correlations and causes the clustering of clusters to increase. 

We combine the evolution of the cluster correlation function 
with its dependence on cluster mass by determining the evolution of 
the richness-dependent cluster correlation function. We do so using 
the format of the correlation scale versus mean separation relation, 
$R_0 - d$. This relation is easy to compare with 
observations. Samples with intrinsically larger mean separations 
correspond to lower intrinsic cluster abundances and thus to higher 
cluster richness and mass (for complete samples). We find that, 
remarkably, the richness-dependent cluster correlation function 
$R_0 - d$ is independent of redshift for these models for $z \la 2$ for 
LCDM (using a fixed correlation function slope and cluster mass defined within a fixed comoving radius) 
and $z \la 1$ for TSCDM (Figure 4). The amplitude of the $R_0 - d$ relation begins to decline  
only at $z \ga 2$ for LCDM and $z \ga 1$ for TSCDM. Using a free correlation slope fit, or virial cluster masses, 
yields similar results but with a small amount of evolution ($\la 15\%$ to $z \la 2$ for LCDM).  

The non-evolving $R_0 - d$ relation implies that the strengthening 
of the cluster correlation function with redshift is matched by 
the relevant increase in the mean separation at high redshift 
(lower cluster abundance). The same mass clusters shift upwards 
along the $R_0 - d$ relation as the redshift increases: both the 
comoving clustering scale and the cluster mean separation increase 
with redshift so that the $R_0 - d$ relation remains essentially unchanged. 
At $z \la 2$, the LCDM relation follows approximately 
$R_0 (z) \simeq 2.6 \sqrt{d(z)}$ (comoving scales).

The evolution of the $R_0 - d$ relation to high redshift provides 
an important new cosmological tool. The observed evolution of the 
relation from $z \simeq 0$ to $z \sim 3$ can be used to break degeneracies 
that exist at $z \sim 0$ and thus allow a precise determination of 
cosmological parameters. No evolution in the $R_0 - d$ relation is  
expected (using a correlation slope $\gamma = 2$) for $z \la 2$ if the current 
LCDM model is correct.

\clearpage


\epsscale{0.92}
\begin{figure}
\plotone{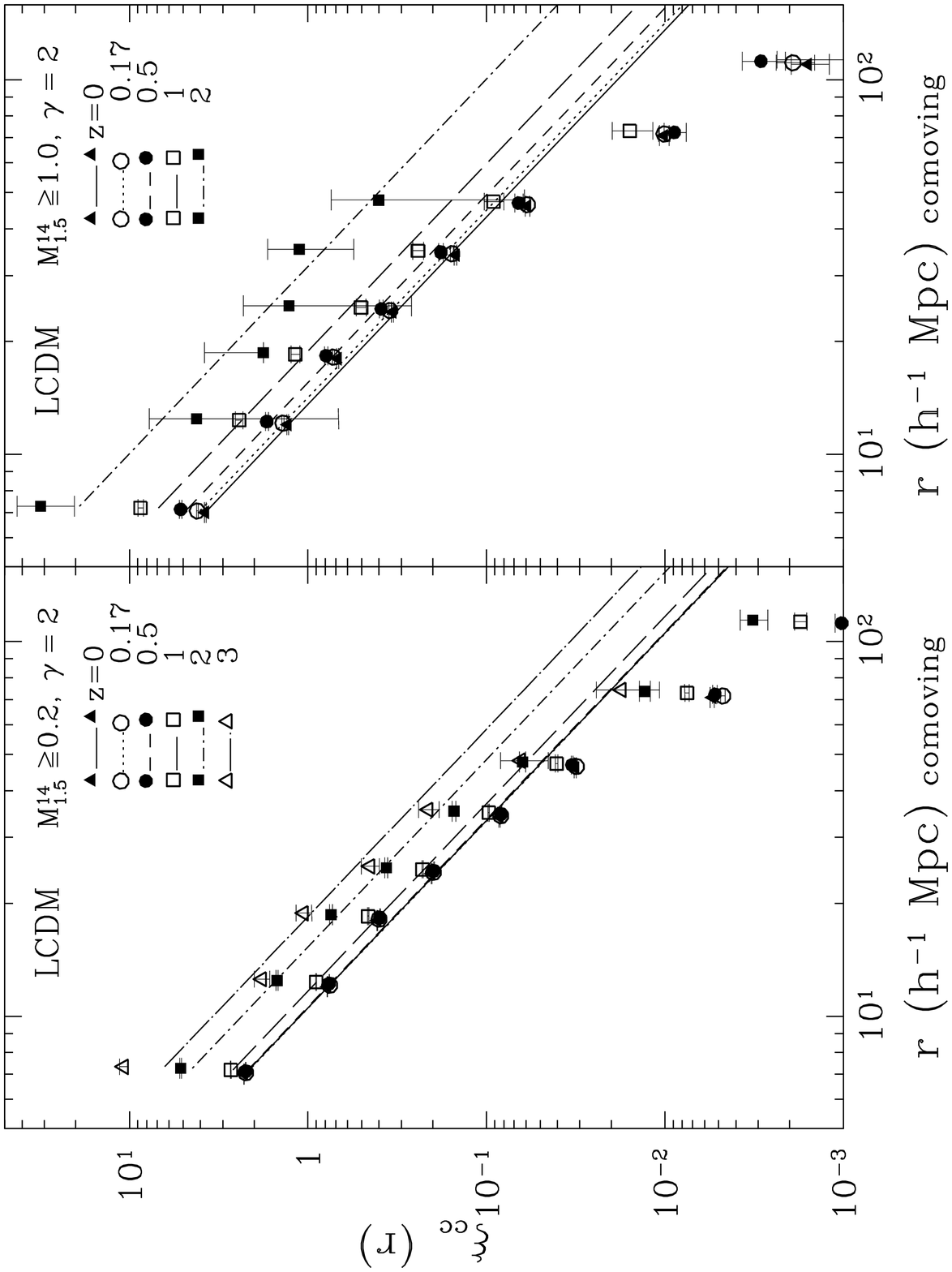}
\caption{The cluster correlation functions at different redshifts,
from $z = 0$ to $z = 3$, for two mass threshold samples ($M$($r\leq 1.5\ h^{-1}$ Mpc 
comoving) $\geq 0.2\times10^{14}$ and $1\times10^{14}\ h^{-1}$ \msun), for the LCDM model.
The lines are the best-fit correlation functions (for $r\leq50 \ h^{-1}$ Mpc) 
for the different redshift samples, using a fixed slope of $\gamma = 2$.
(1-$\sigma$ Poisson error-bars).
\label{f1}}
\end{figure}

\epsscale{0.92}
\begin{figure}
\plotone{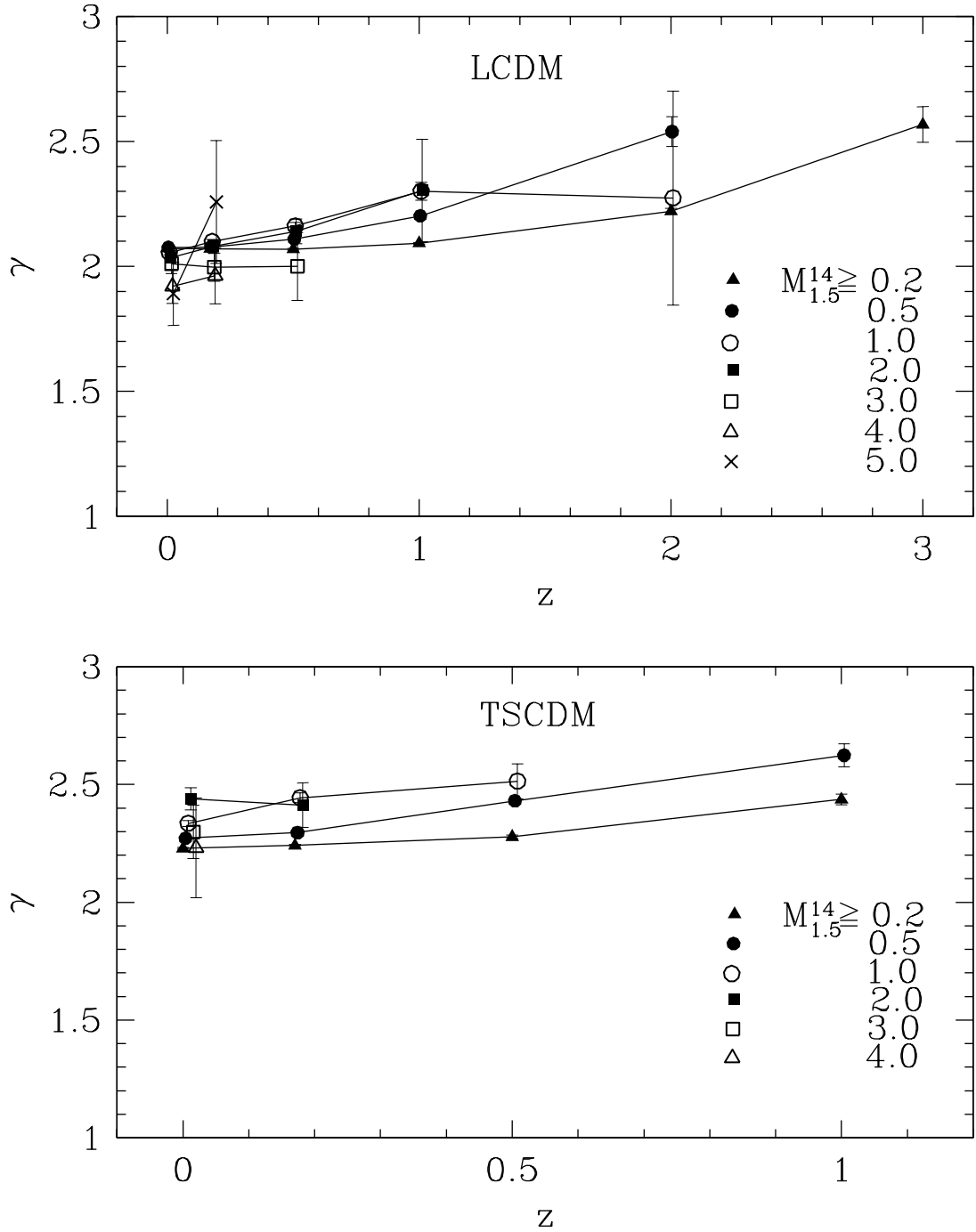}
\caption{The best-fit correlation function slope (for $r\leq 50 h^{-1}$ Mpc) as a function
of redshift and cluster mass threshold for LCDM and TSCDM models.
\label{f2}}
\end{figure}

\epsscale{0.92}
\begin{figure}
\plotone{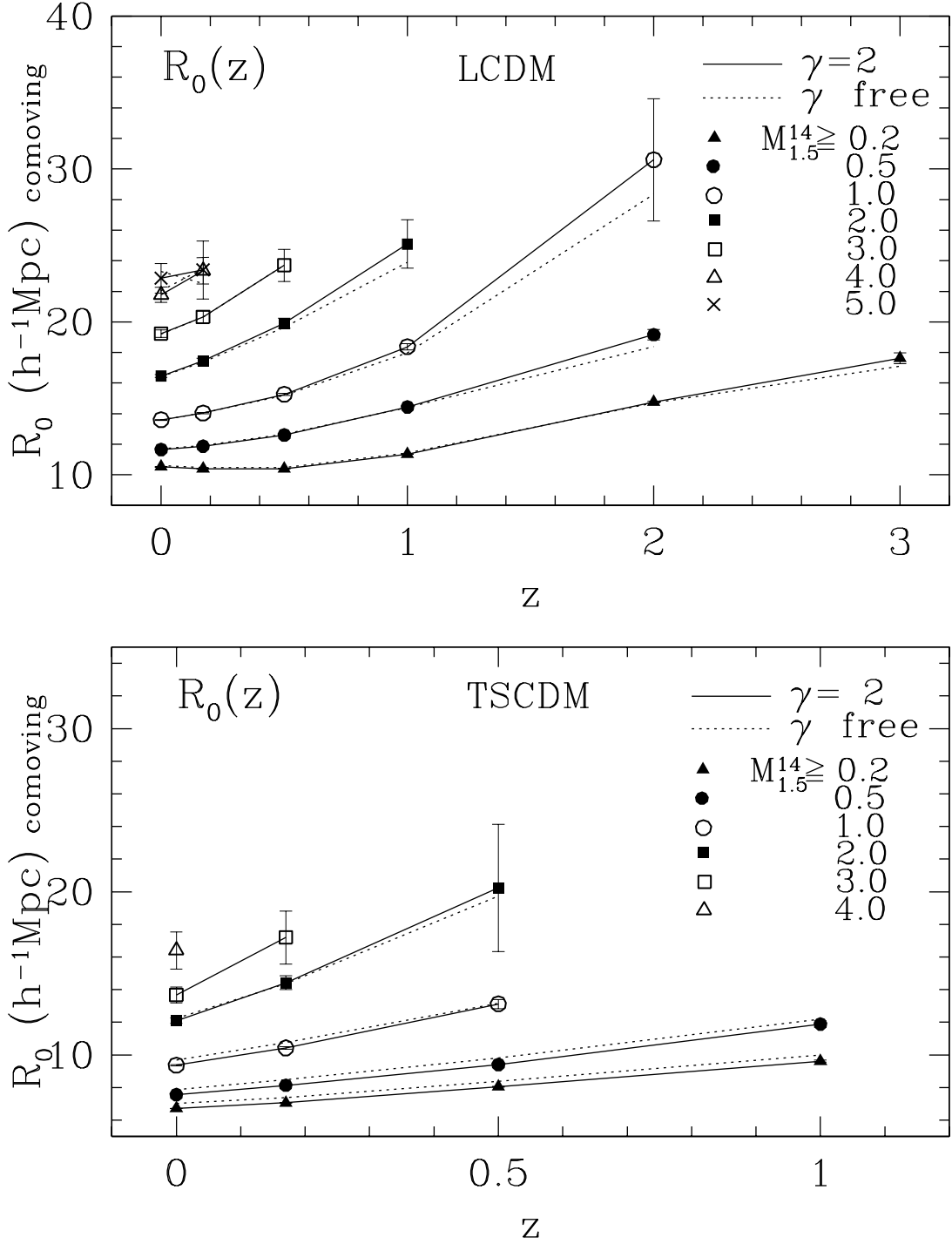}
\caption{The best-fit cluster correlation scale ($R_0$, comoving units)
as a function of redshift and cluster mass ($M_{1.5}$). Symbols and 
connected solid lines are for a fixed correlation slope of $\gamma = 2$;
dotted lines are for the free-fit correlation slope (Figure 2). 
\label{f3}}
\end{figure}

\epsscale{0.75}
\begin{figure}
\plotone{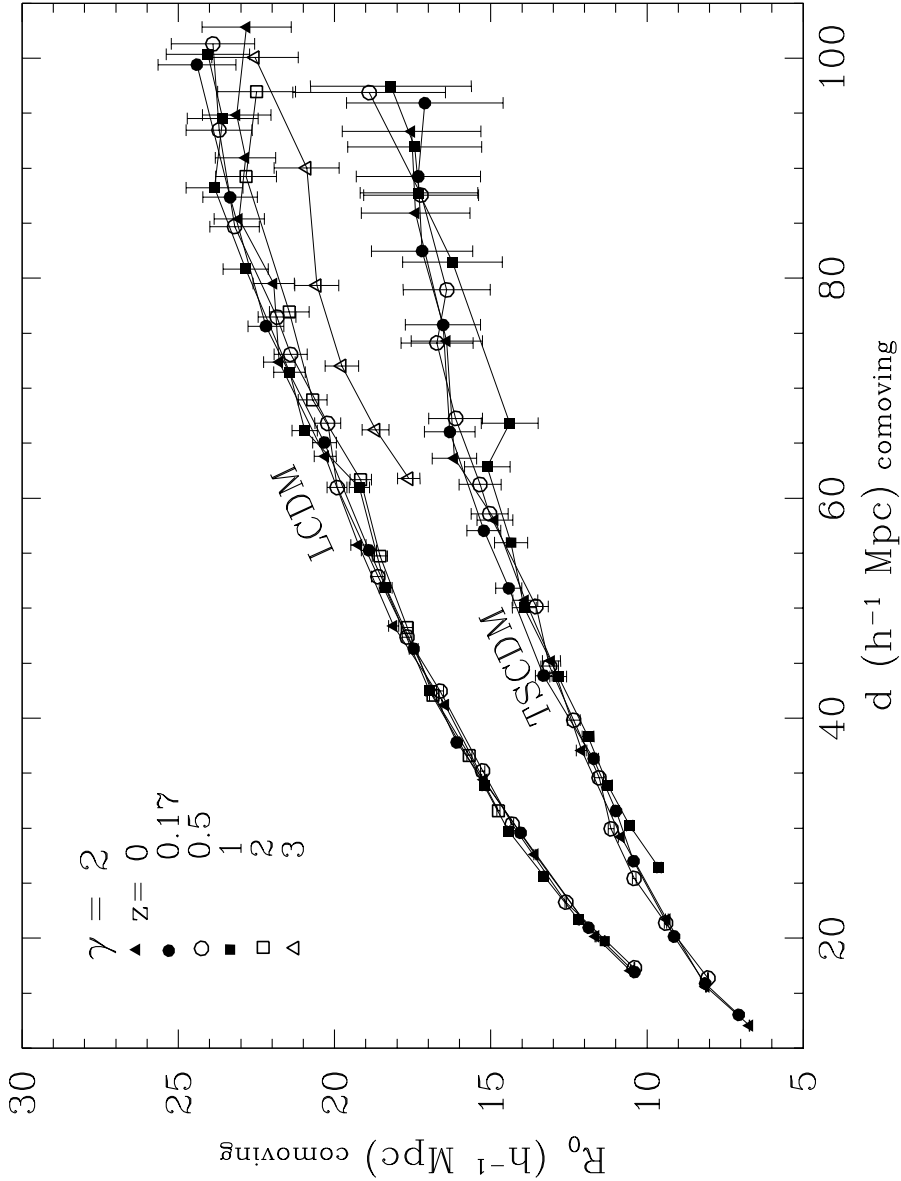}
\caption{Evolution of the richness-dependent cluster correlation 
function, presented as the dependence of the correlation scale on 
the cluster mean-separation, $R_0 - d$ (comoving scales). Different 
symbols represent different redshifts, from $z = 0$ to $z = 3$, as indicated. 
The lines connect points of a given redshift. The evolution of the 
$R_0 - d$ relation is shown for both LCDM and TSCDM. A fixed 
correlation function slope of $\gamma = 2$ is used, and cluster masses are defined 
within a fixed comoving radius (1.5 $h^{-1}$ Mpc). (Using clusters 
selected instead by virial mass, which is considerably more difficult 
to determine observationally, produces similar results but with 
slightly more evolution at high $d$'s: the $R_0 - d$ amplitude decreases 
by $\la 10\%$ as the redshift increases from $z = 0$ to $z = 2$ for LCDM. The 
$R_0$ amplitude is lower by $\la 10\%$ at small $d \ \la40\ h^{-1}$ Mpc for these masses; 
see Section 4). 1-$\sigma$ error-bars are shown. 
\label{f4}}
\end{figure}

\epsscale{0.92}
\begin{figure}
\plotone{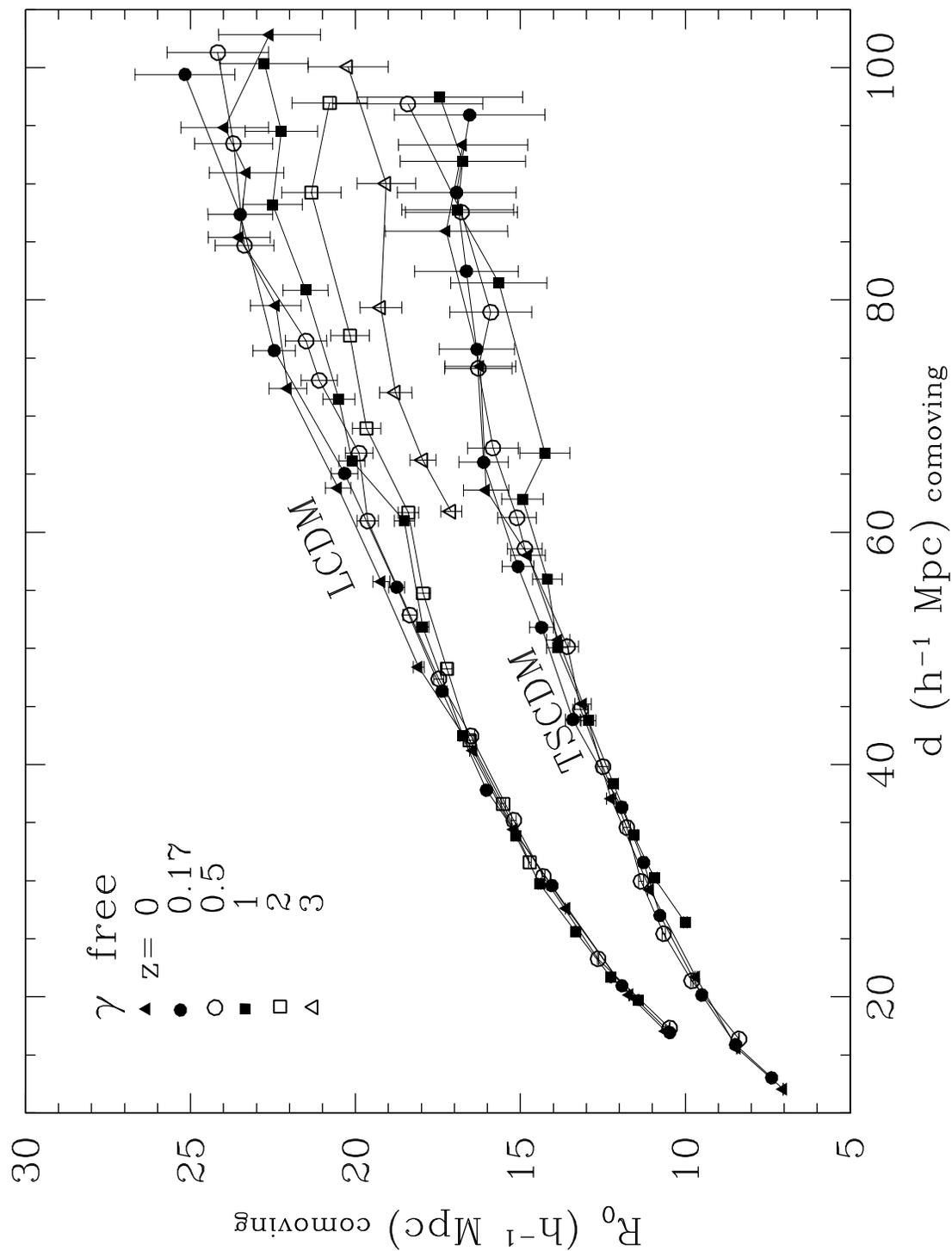}
\caption{
Evolution of the richness-dependent cluster correlation
function. Same as Figure 4, but for a free-fit correlation function 
slope $\gamma$ (c.f. Figures 2-3).
\label{f5}}
\end{figure}

\clearpage


\end{document}